\documentclass{article}
\usepackage{amsmath,amssymb,amsfonts,url}
\usepackage{graphicx}
\usepackage{color}
\usepackage{subfig}
\usepackage[margin=1in]{geometry}

\title{Early MFCC And HPCP Fusion for Robust Cover Song Identification}
\date{}

\author
 {Christopher J. Tralie \\ Duke University Department of Electrical and Computer Engineering \\ \texttt{ctralie@alumni.princeton.edu }}

\begin{document}

\maketitle
\begin{abstract}

While most schemes for automatic cover song identification have focused on note-based features such as HPCP and chord profiles, a few recent papers surprisingly showed that local self-similarities of MFCC-based features also have classification power for this task.  Since MFCC and HPCP capture complementary information, we design an unsupervised algorithm that combines normalized, beat-synchronous blocks of these features using cross-similarity fusion before attempting to locally align a pair of songs.  As an added bonus, our scheme naturally incorporates structural information in each song to fill in alignment gaps where both feature sets fail.  We show a striking jump in performance over MFCC and HPCP alone, achieving a state of the art mean reciprocal rank of 0.87 on the Covers80 dataset.  We also introduce a new medium-sized hand designed benchmark dataset called ``Covers 1000,'' which consists of 395 cliques of cover songs for a total of 1000 songs, and we show that our algorithm achieves an MRR of 0.9 on this dataset for the first correctly identified song in a clique.  We provide the precomputed HPCP and MFCC features, as well as beat intervals, for all songs in the Covers 1000 dataset for use in further research.

\end{abstract}

\section{Introduction}
A ``cover song'' is a different version of the same song, usually performed by a different artist, and often with different instruments, recording settings, mixing/balance, tempo, and key.  To sidestep a rigorous definition, like others, we evaluate algorithms on a set of songs that have been labeled as covers of each other, and we declare success when our algorithm recognizes clusters of songs which have been deemed covers of each other.  In fact, this problem is more of a ``high level music similarity'' task beyond exact recording retrieval, making the problem intrinsically more difficult than traditional audio fingerprinting \cite{haitsma2002highly, wang2003industrial, wang2006shazam}.

Most work on automatic cover song identification to date has focused on estimating and matching note-based features such as chord estimates \cite{bello2007audio}, chroma \cite{ellis2006identifying, ellis2007}, harmonic pitch class profiles (HPCP) \cite{gomez2006song, serra2007music,  serra2008chroma}, 2D Fourier magnitude coefficients to approximate these features \cite{ellis2012large, nieto2014music}.  This is natural, since regardless of all of the transformations that can happen between versions, note sequences should be preserved up to transpositions.  Problems occur, however, when note sequences are not the defining characteristic of a musical expression.  This is common in hip hop, for example, such as the song ``Tricky'' in the ``Covers 80 Dataset'' (\cite{ellis2007covers80}, Section~\ref{sec:covers80}) performed by Run D.M.C. and The Beastie Boys.  There are also songs which are entirely percussive, such as the 8 covers of Frank Zappa's song ``The Black Page'' that we present in Section~\ref{sec:zappa}.  Moreover, even for song pairs with strong harmonic content, there may be sections with drum solos, un-pitched spoken words, or other musical statements on which pitch-based features fail.  However, the issue with features complementary to Chroma, such as Mel-Frequency Cepstral Coefficients (MFCCs), is that they are highly sensitive to instrument and balance changes.  In spite of this, some recent works have shown that MFCC-based features can also be used in cover song identification \cite{chen2015modified, tralie2015cover}.  Particularly, if the {\em relative} changes of MFCC are captured, as in \cite{tralie2015cover}, performance is still reasonable.


Naturally, then, rather than relying on a single feature in isolation, recent works have shown the benefits of feature fusion after song comparisons have been made with each feature set alone.  For instance, aggregating ranks from individual features can improve results \cite{osmalsky2015combining,osmalsky2016enhancing}.  Other works show the advantage of using all pairwise similarities computed with different features \cite{Chen2017CSFusion}, using a cross-diffusion process known as ``similarity network fusion'' (SNF) (\cite{wang2012unsupervised, wang2014similarity}, Section~\ref{sec:snfdef}) to come up with a consensus similarity score between all pairs of songs in a corpus.

In this work, we develop a similarity network-based early fusion technique which achieves state of the art results by combining complementary HPCP, MFCC, and self-similarity MFCCs (SSM MFCCs).  Unlike \cite{Chen2017CSFusion}, we apply SNF {\em before alignment} in the space of features.  This fusion technique incorporates both cross-similarity between two different songs and self-similarity of each song, so it is able to combine information about matching sections between songs and structural elements within each song.  We also apply late fusion on similarity scores between a network of songs to further boost the results.  While state of the art supervised techniques on the popular ``Covers 80'' benchmark dataset yield a mean reciprocal rank (MRR) of 0.625 \cite{Chen2017CSFusion}\footnote{This technique scored the best in MIREX 2016}, our completely unsupervised technique achieves a MRR of 0.87 (Section~\ref{sec:covers80}).  We also introduce our own dataset consisting of 395 cliques of songs, which we call ``Covers1000'' (Section~\ref{sec:covers1000}), and on which we report an MRR of 0.9 with our best fusion technique.

\section{Beat-Synchronous Blocked Features}
\label{sec:blockpc}

In this section, we describe three complementary features which we later fuse together in Section~\ref{sec:coverfusion}.  One concept we rely on in our feature design is ``block-windowing,'' which was described in \cite{tralie2015cover}.  The idea is that when a contiguous set of features in time are stacked into one large vector, they give more information about the local dynamics of a song.  This is related to the concept of a delay embedding in dynamical systems \cite{kantz2004nonlinear}.  Having many blocks across the song also allows us to control for drift by normalizing within each block.  To control for tempo differences, we compute our blocks synchronized with beats, which is a common preprocessing step \cite{ellis2006identifying, ellis2007, tralie2015cover}.  We use the simple dynamic programming approach of \cite{ellis2007beat}, since it allows the specification of a tempo bias.  As in \cite{ellis2007} and \cite{tralie2015cover}, we bias the beat tracker at 3 different tempo levels (60, 120, 180bpm), and we compare all pairs of tempo levels, for up to 9 unique combinations, since the beat tracker may choose an arbitrary level of subdivision in a rhythm hierarchy.  Once the beat onsets are computed, we form a block for every contiguous group of $B$ beat intervals, as in \cite{tralie2015cover}.

\subsection{HPCP Features}

One proven set of pitch-based features for cover song identification are ``harmonic pitch class profiles'' (HPCPs) \cite{gomez2006tonal} \footnote{To ensure we have a state of the art implementation, we use the Essentia library to compute HPCPs \cite{bogdanov2013essentia}.}.  Following \cite{serra2009cross}, we compute a stacked delay embedding of the HPCP features within $B$ beats, with two HPCP windows per beat, for a total of $2B$ windows per block.  This has an advantage over other works which do not use a delay, as it gives more context in time, and it is consistent with the block/windowing framework.  To normalize for key transpositions, we need to determine an ``optimal transposition index'' (OTI) between two songs (\cite{serra2008transposing}).  Given the average HPCP vector $X \in \mathbb{R^+}^{12}$ from song A and the average HPCP vector $Y \in \mathbb{R^+}^{12}$ from song B, we compute the correlation $X^T Y$ over all 12 half-step transpositions of the original HPCP features in the block, and we use the transposition that leads to the maximum correlation.  Then, we compute cosine distance between all pairs of HPCP blocks between the two songs.


\subsection{MFCCs / MFCC Self-Similarity Matrices (SSMs)}
\label{sec:SSMs}

\begin{figure}
    \centering
    \includegraphics[width=0.6\textwidth]{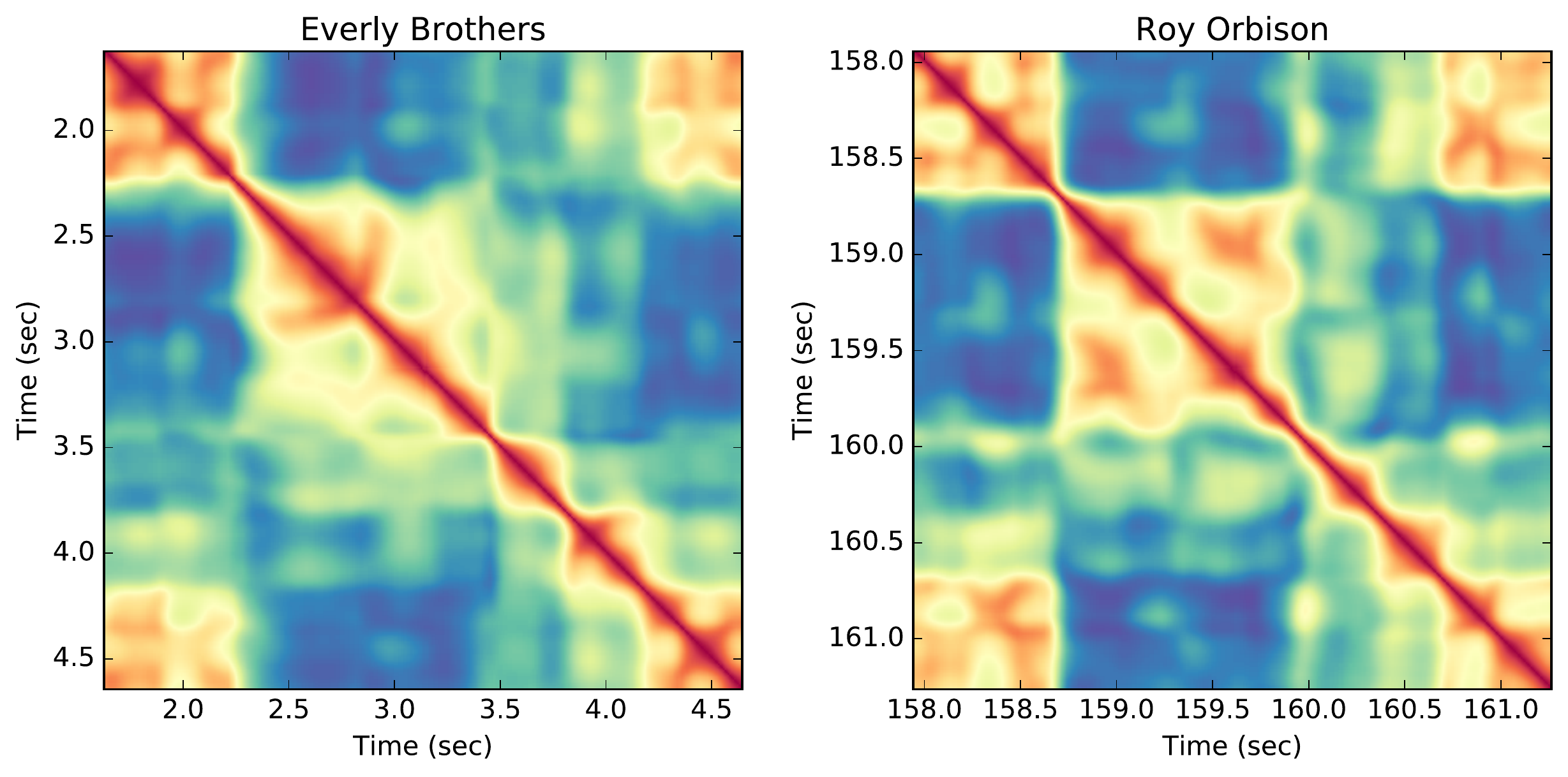}
    \caption{Example 8-beat Z-normalized MFCC SSMs blocks in correspondence between cover versions.  A block from ``Claudette'' by the Everly Brothers and Roy Orbison.  The pattern in this block is Guitar + ``Oooh ooh Claudette'' + Guitar.}
	\label{fig:MFCCSSMExamples}
\end{figure}

In addition to HPCP features, we compute exponentially liftered MFCCs in beat-synchronous blocks.  We take the MFCC window size to be 0.5 seconds, and we advance the window intervals evenly from the beginning of the block to the end of a block with a {\em hop size} of $512$ samples.  At a sample rate of $44100$ Hz, this leads to a window size of $22050$ and an overlap of roughly $97.5 \%$ between windows (Section~\ref{sec:SSMs}).  Longer windows have been shown to increase robustness of SSM matching in \cite{tralie2015cover} and audio fingerprinting \cite{haitsma2002highly}, which justifies this choice.  To allow direct comparisons between different blocks, we interpolate to 400 MFCCs per block, and we perform Z-normalization (as in \cite{tralie2015cover}) to control for loudness and drift, which we found to be an essential step.

In addition to block-synchronized and normalized raw MFCCs, we also compute self-similarity matrices (SSMs) of the Z-normalized MFCCs within each block, as in \cite{tralie2015cover}, leading to a sequence of SSMs for each song.  That is, unlike \cite{bello2009SSMStructure}, who compares SSMs between entire songs, we compare SSMs summarizing blocks of audio on the order of tens of beats, as recommended by \cite{tralie2015cover}.  For each beat-synchronous block, we create a Euclidean SSM between all MFCC windows in that block.  As with the raw MFCCs, to allow comparisons between blocks, we resize each SSM to a common image dimension $d \times d$.  Figure~\ref{fig:MFCCSSMExamples} shows two examples of MFCC SSM blocks with 8 beats and 500 windows per block which were matched between a song and its cover in the Covers80 dataset.  Although the underlying sounds are quite different (male to female, different instruments and balance), the SSMs look similar.  \cite{tralie2015cover} argue that this is why, counter to prior intuition, it is possible to use MFCCs in cover songs.

\subsection{Cross-Similarity Matrices (CSMs) between Blocks}
\label{sec:globalcompare}

Given a set of $M$ beat-synchronous block features for a song A and a set of $N$ beat-synchronous block features for a song B, we compare all pairs of blocks between the two songs for that feature set, yielding an $M \times N$ cross-similarity matrix (CSM), which can be used to align the songs.  For each song, we have 3 different CSMs for the three different feature sets, which are each aligned at the same beat intervals.  For MFCC and MFCC SSMs, we use the Euclidean distance (Frobenius norm) to create the CSM, while for HPCP, we use the cosine distance after OTI.  We then use the Smith Waterman algorithm \cite{smith1981identification} to find the best locally aligned sections between a pair of songs.  To apply Smith Waterman, we turn the CSM into a binary matrix $B^M$, so that $B^M_{ij} = 1$ if $CSM_{ij}$ is within the $\kappa N^{\text{th}}$ smallest values in row $i$ of the CSM and if $CSM_{ij}$ is within the $\kappa M^{\text{th}}$ smallest values in column $j$ of the CSM, and 0 otherwise, where $\kappa \in (0, 1)$ is a {\em mutual nearest neighbors} threshold.  As shown by \cite{serra2008chroma} and \cite{tralie2015cover}, this is a crucial step for improving robustness.


Once we have the $B^M$ matrix, we use a diagonally constrained variant of Smith Waterman to locally align the two songs.  The score returned by this algorithm roughly corresponds to the length of the longest interval of consecutive blocks matched between songs, with more tolerance for gaps than a naive cross-correlation.  More details can be found in\cite{serra2008chroma} and \cite{tralie2015cover}.  For comparison, we perform alignments with the CSMs obtained for each feature individually and on the CSM obtained from fusing them.

\section{Feature Fusion}
\label{sec:coverfusion}

\begin{figure*}[h]
	\centering
	\includegraphics[width=\textwidth]{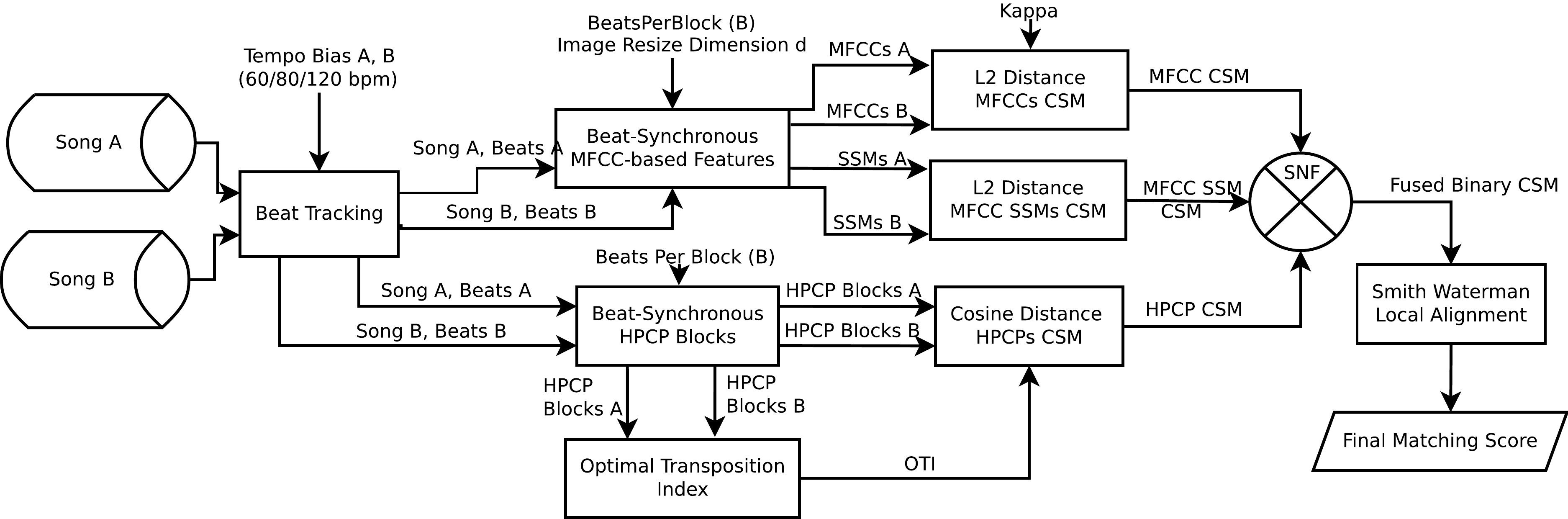}%
	\caption{A block diagram of our system which performs early similarity network fusion of blocked MFCCs, MFCC SSMs, and HPCPs before scoring with Smith Waterman alignment.}
	\label{fig:BlockDGMFusion}
\end{figure*}

Figure~\ref{fig:BlockDGMFusion} shows an overall pipeline for the fusion process.  We first briefly review the general mathematical technique that we use to fuse cross-similarity matrices from different feature sets, and then we show specifically how we apply it in our pipeline.

\subsection{Similarity Network Fusion (SNF)}
\label{sec:snfdef}


Given all pairs of similarity scores between objects using different features, Similarity Network Fusion (SNF) is designed to find a better matrix of all pairwise scores that combines the strengths of each feature \cite{wang2012unsupervised, wang2014similarity}.  Roughly, it performs a random walk using the nearest neighbor sets from one feature while using the transition probability matrices from all of the other features, while continuously switching which feature set is used to determine the nearest neighbors.  More precisely, the algorithm is defined as follows.  First, start with a pairwise distance function $\rho(i, j)$ between objects for each feature type, and create an exponential kernel $W(i, j) = e^{-\rho^2(i, j)/2(\sigma_{ij})^2}$, where $\sigma_{ij}$ is a neighborhood-autotuned scale which is the average of $\rho(i, j)$ and the mean $k$-nearest neighbor intervals of blocks $i$ and $j$, for some $k$ (see \cite{wang2014similarity} for more details).  Now, create the following Markov transition matrices

\begin{equation}
P(i, j) = \left\{  \begin{array}{cc} \frac{1}{2} \frac{W(i, j)}{\sum_{k \neq i} W(i, k)} & j \neq i \\ 1/2 & \text{otherwise} \end{array} \right\}
\end{equation}

This is simply a first order Markov chain with a regularized diagonal to promote self-recurrence.  Once this matrix has been obtained, create a truncated $k$-nearest neighbor version of this matrix

\begin{equation}
S(i, j) = \left\{  \begin{array}{cc} \frac{W(i, j)}{\sum_{k \in N(i)} W(i, k)} & j \in N(i) \\ 0 & \text{otherwise} \end{array} \right\}
\end{equation}

where $N(i)$ are the $k$ nearest neighbors of vertex $i$, for some chosen $k$.  Now let $P^f$ and $S^f$ be the $P$ and $S$ matrices for the $f^{\text{th}}$ feature set, and let $P^f_{t = 0} = P^f$.  Then define a first order ``cross-diffusion'' random walk recursively as follows

\begin{equation}
P_{t+1}^f = S^f \left( \frac{\sum_{v \neq f} P^v_t}{m-1} \right)(S^f)^T
\end{equation}

where $m$ is the total number of features.  In other words, a random walk is occurring but with probabilities that are modulated by similarity kernels from other features.  As shown by \cite{wang2012unsupervised}, this process will eventually converge, but we can cut it off early.  Whenever it stops, the final fused transition probabilities are $\hat{P}_t = \frac{1}{m} \sum_{k = 1}^M P_t^k$.

\subsection{Late SNF}
\label{sec:latesnf}

One way to use SNF is to let the matrix $\rho(i, j)$ be all pairwise distance between songs, computed by some alignment \cite{Chen2017CSFusion}.  In this case, the result should be a better network of similarity scores between all songs \footnote{Note that \cite{serra2012characterization} essentially do the same thing with only one feature set}.  We follow a similar approach to \cite{Chen2017CSFusion}, but we we work with the Smith Waterman scores we get from a unique combination of MFCC, MFCC SSM, and HPCP blocks (\cite{Chen2017CSFusion} applied SNF to different alignment schemes on the same feature set).  Given a particular score matrix $S$ between all pairs of songs, we compute the kernel matrix $W$ as $W(i, j) = 1/S(i, j)$.  Since Smith Waterman gives a higher score for better matching songs, this ensures that the kernel is close to $0$ in this case.  At this point, we perform SNF, and we obtain a final $N \times N$ transition probability matrix $P$.  We can then look along each row to find the neighboring songs with maximum fused probability.  This process can be thought of as exploiting the {\em network} of all songs in a collection in an unsupervised manner.

\subsection{Early SNF}
\label{sec:earlysnf}

\begin{figure}
	\centering
	\includegraphics[width=0.35\textwidth]{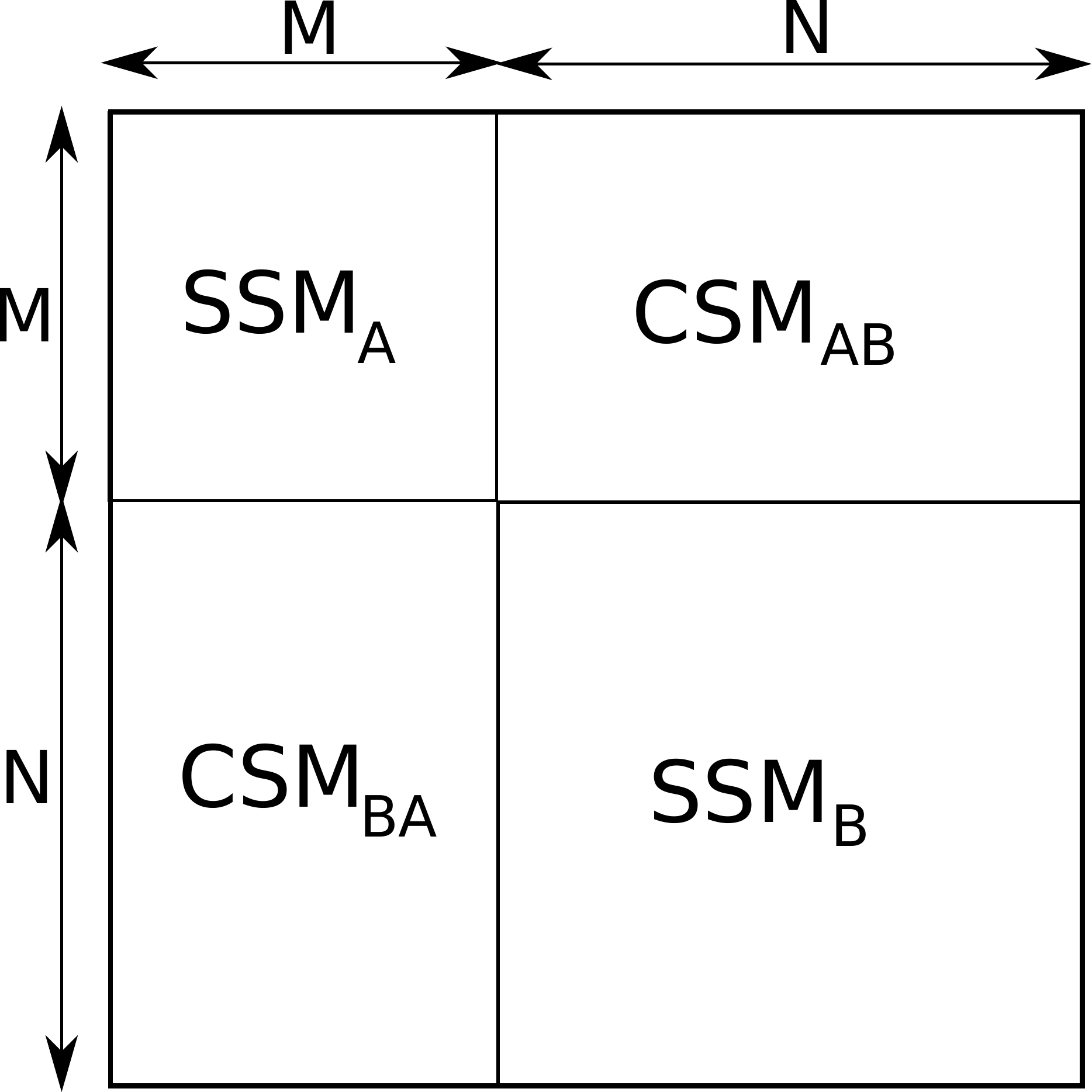}
	\caption{A pictorial representation of the SSM that results when concatenating song $B$ to song $A$, which we feed to SNF for early fusion of low level features.  Including self-similarity blocks of each song helps to promote structural elements in cross-similarity regions during SNF.}
	\label{fig:ParentSSM}
\end{figure}


In addition to SNF after Smith Waterman has given scores, we can perform fusion at the feature level before running Smith Waterman.  One advantage of doing fusion before scores are computed is that we don't need a network of songs to compute a score; we can obtain an improved score between two songs without any other context\footnote{Note that \cite{Chen2017CSFusion} refer to SNF after Smith Waterman as ``early fusion'' with respect to rank aggregation, which they call ``late fusion,'' but we call their technique ``late fusion'' because we fuse before Smith Waterman with SNF, which is even earlier in the pipeline.}.  Our technique for early fusion, which we found to be superior to the ``OR fusion'' proposed in \cite{foucard2010multimodal}, is to apply SNF on the cross-similarity matrices obtained from two or more different feature sets before creating a binary CSM and sending it off to Smith Waterman.

As defined in Section~\ref{sec:snfdef}, SNF operates on self-simlarity matrices (SSMs), so it cannot be directly applied to this problem.  To make it so that CSMs fit into the framework, we create a ``parent SSM'' for each feature set that holds both SSMs and the CSM for that feature set.  In particular, given song $A$ with $M$ blocks in a particular feature set and song $B$ with $N$ blocks in that feature set, form the SSM $D_{AB}$ which is the SSM that results after concatenating song $B$ to the end of song $A$.  Let the SSM for song $A$ be $D_A$, the SSM for song $B$ be $D_B$, and the CSM between them be $C_{AB}$.  Then $D_{AB}$ can be split into four sub-blocks:

{
\footnotesize
\begin{equation}
D_{AB}(i, j) = \left\{ \begin{array}{cc}
	D_A(i, j) & i < M, j < M  \\
	D_B(i-M, j-M) & i >= M, j >= M \\
	C_{AB}(i, j-M) & i < M, j >= M \\
	C_{BA}(i-M, j) =  &  \\
    C_{AB}^T(j, i-M) & i >= M, j < M
	\end{array}\right\}
\end{equation}
}

\begin{figure}
	\centering
	\includegraphics[width=0.9\textwidth]{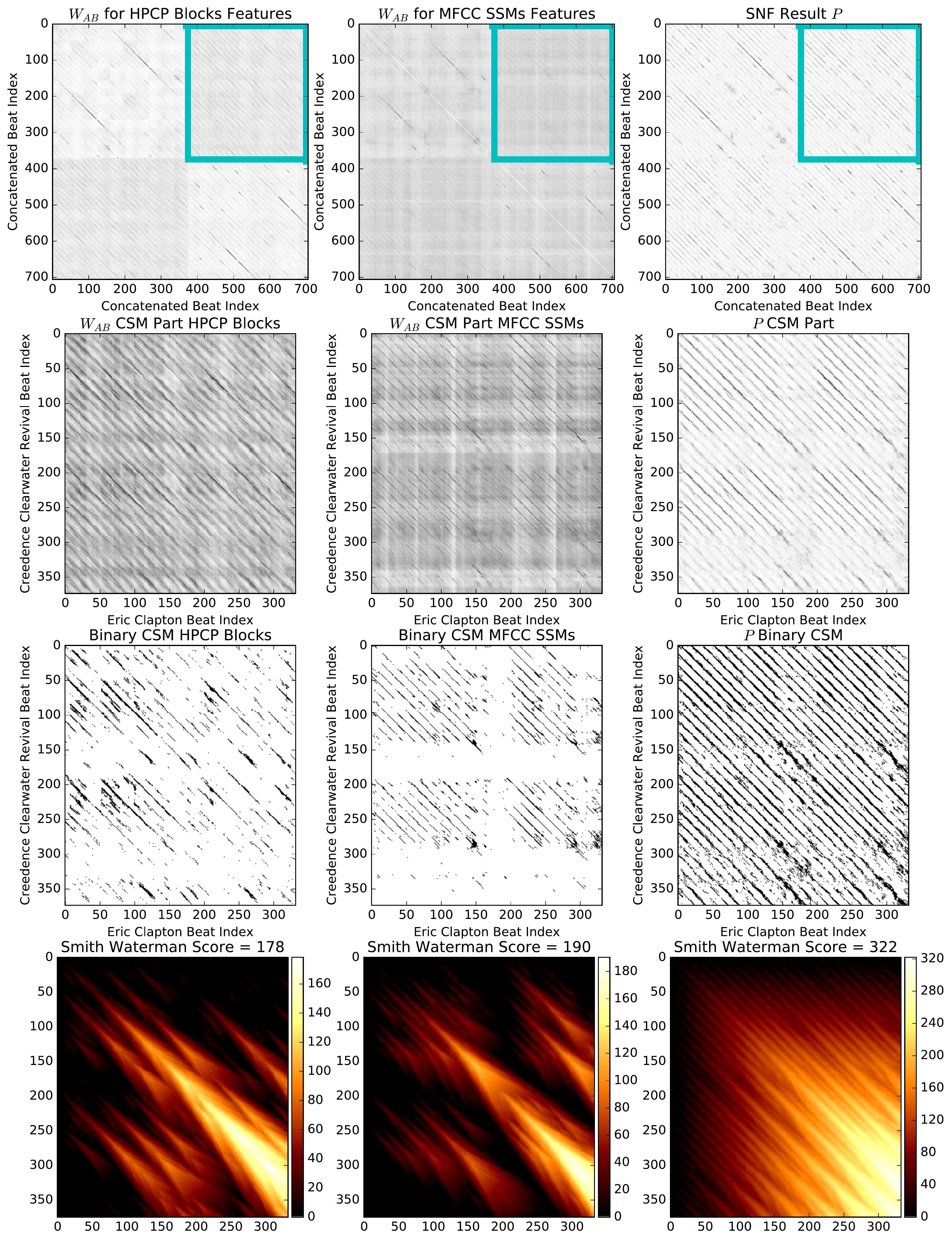}
	\caption{An example of early SNF on blocks of MFCC SSMs and blocks of HPCP features on the song ``Before You Accuse Me'' with versions by Eric Clapton and Creedence Clearwater Revival.  The block size is $20$ beats, and there are three iterations of SNF. The kernels $W_{AB}$ are shown for each, and the CSM portion is highlighted with a blue box.  The final fused probability matrix $P$ returned from SNF is shown in the upper right.  The corresponding CSM portions for all three matrices shown for each on the bottom.  In the fused probability matrix, the diagonal regions are much crisper and more distinct from the background than they are for the individual feature sets.  The result is that the mutual nearest neighbors binary CSM has longer uninterrupted diagonals, which is reflected by a higher Smith Waterman score.}
	\label{fig:EarlySNFExample}
\end{figure}

Figure~\ref{fig:ParentSSM} shows this pictorially. Given such a matrix for each feature set, we could then run SNF and extract the cross-similarity sub-matrix at the end.  The issue with this is the dynamic range of the SSM may be quite different from the dynamic range of the CSM, as it is likely that blocks in song $A$ are much more similar to other blocks in song $A$ than they are to blocks in $B$.  To mitigate this, given a nearest neighbor threshold $\kappa$ for the CSM, we compute the kernel neighborhood scales $\sigma_{ij}$ individually for $D_A$, $D_B$, and $CSM_{AB}$, and we put them together in the final kernel matrix $W_{AB}$ according to Figure~\ref{fig:ParentSSM}.  Once we have such a matrix $W_{AB}$ for each feature set, we can finally perform SNF.  At the end, we will end up with a fused probability matrix $P$, from which we can extract the cross-probability $P_{C_{AB}}$.  We can then take mutual highest probabilities (akin to mutual nearest neighbors) to extract a binary matrix and perform Smith Waterman as normal.  Figure~\ref{fig:EarlySNFExample} shows an example of constructed matrices $W_{AB}$ and the resulting fused probabilities $P$.

One advantage of this technique is that since the CSM and SSMs are treated together and normalized to a similar range, any recurrent structure which exists in the SSMs can reinforce otherwise weaker structure in the CSMs during the diffusion process.  This can potentially help to strengthen weaker beat matches in an otherwise well-matching section, leading to longer uninterrupted diagonals in the resulting binary CSM.

\subsection{Early Fusion Examples}

\begin{figure*}
	\centering
    \includegraphics[width=0.8\textwidth]{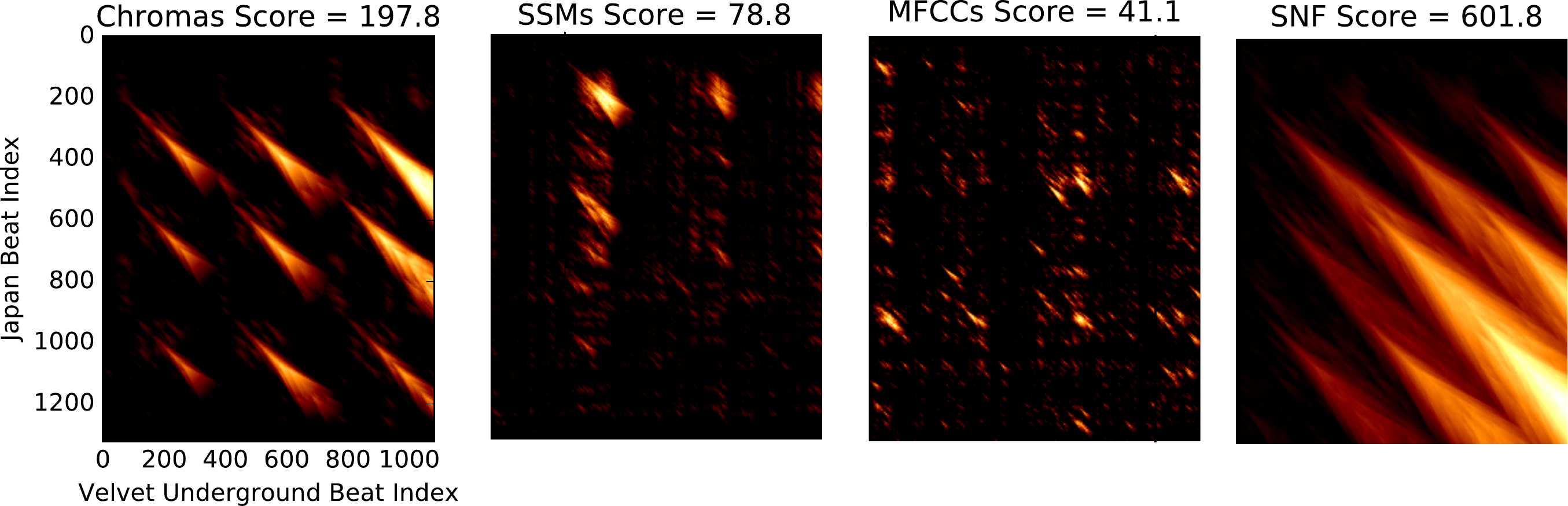}
	\caption{Smith Waterman tables/scores for ``All Tomorrow's Parties'' by Japan and Velvet Underground.}
	\label{fig:AllTomorrowsParties}
\end{figure*}

\begin{figure*}
	\centering
	\includegraphics[width=0.9\textwidth]{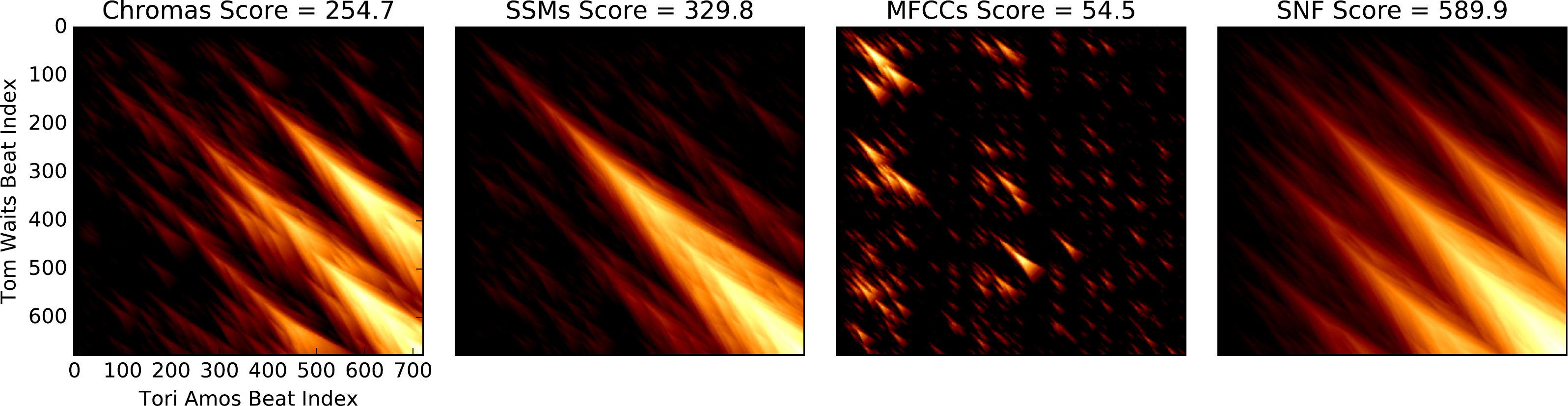}
	\caption{Smith Waterman tables/scores for ``Time'' by Tom Waits and Tori Amos}
	\label{fig:TimeTheSong}
\end{figure*}

Before we launch into a more comprehensive experiment, we show a few examples of early SNF to illustrate the value added.   In each example, we used 20-beat blocks, $\kappa = 0.1$ for both similarity fusion and binary nearest neighbors, and 3 iterations of SNF.  Figure~\ref{fig:AllTomorrowsParties} shows an example where the three individual features are rather poor by themselves, but where they happen to all pick up on similarities in complementary regions.  As a result, early SNF does a fantastic job fusing the features.  Figure~\ref{fig:TimeTheSong} shows an example where MFCC SSMs happen to do better than HPCP, but where the results fusing both are still better than each individually.

\section{Experiments}
\label{sec:Results}

We are now ready to evaluate the performance of this new algorithm.  In all of our experiments below, we settle on $\kappa = 0.1$ (the mutual nearest neighbor threshold for binary CSMs) and $B = 20$ beats per block.  For similarity network fusion, we take 20 nearest neighbors for both early fusion and late fusion, we perform 3 iterations for early fusion, and we perform 20 iterations for late fusion.  We also include an ``early + late'' fusion result, which is applying late fusion to the network of similarities obtained from all of the feature sets (MFCCs, MFCC SSMs, HPCPs) plus the network of similarities obtained from the early fusion of the three feature sets.

\subsection{Covers 80 Dataset}
\label{sec:covers80}

\begin{table}
\centering
\caption{Results of different features and fusion techniques on the Covers 80 dataset.}
\begin{tabular}{|c|c|c|c|c|c|}
\hline
 & MR & MRR & Top-01 & Top-10 & ../80 \\ \hline
MFCCs & 29.7 & 0.538 & 79 & 97 & 42/80 \\ \hline
SSMs & 15.1 & 0.615 & 91 & 111 & 48/80 \\ \hline
HPCPs & 18.2 & 0.673 & 102 & 119 & 53/80 \\ \hline
\begin{tabular}[x]{@{}c@{}}Late\\SSMs/MFCCs\end{tabular}  & 14.0 & 0.7 & 107 & 125 & 55/80 \\ \hline
Late All & 8.63 & 0.824 & 127 & 141 & 64/80 \\ \hline
Early & 7.76 & 0.846 & 131 & 143 & 68/80 \\ \hline
Early + Late & 7.59 & 0.873 & 136 & 144 & 69/80 \\ \hline
\cite{Chen2017CSFusion} & ? & 0.625 & ?  & 114 &  ? \\ \hline

\end{tabular}
\label{tab:covers80}
\end{table}

To benchmark our algorithm, we begin by testing it on the ``Covers 80'' dataset \cite{ellis2007covers80}.  This dataset contains 160 songs which are split into two disjoint subsets $A$ and $B$, each with exactly one version of a pair of songs, for a total of 80 pairs.  \cite{ellis2006identifying} and \cite{ellis2007} assess performance as follows: given a song in group A, declare its cover song to be the top ranked song in set B, and record the total number of top ranked songs that are correct.  To get a better idea of the performance, we also compute the mean rank (MR), mean reciprocal rank (MRR), and the number of songs correctly identified past a certain number.  All of these statistics are computed on the full set of 160 songs, which is more difficult than simply looking in set $A$ or set $B$.

Table~\ref{tab:covers80} shows the results.  By themselves, HPCP features perform better than MFCC-based features, which is consistent with findings in the literature.  However, there are big improvements when fusing them all.  Surprisingly, we obtain a score of 42/80 just by blocking and normalizing the MFCCs.  This shows the power of having stacked delay MFCCs and of normalizing within each block to cut down on drift.  Also, when fusing MFCCs and MFCC SSMs with late fusion, we get a large performance boost over either alone, showing that SSMs are adding complementary information to the MFCCs they summarize.

\subsection{Covers 1000 Dataset }
\label{sec:covers1000}

\begin{table}
\centering
\caption{Results of different features and fusion techniques on the Covers 1000 dataset.}
\begin{tabular}{|c|c|c|c|c|}
\hline
\textbf{} & MR & MRR & Top-01 & Top-10 \\ \hline
MFCCs & 83.3 & 0.618 & 583 & 679 \\ \hline
SSMs & 72.5 & 0.623 & 581 & 698 \\ \hline
HPCPs & 44.4 & 0.757 & 727 & 809 \\ \hline
Late & 19.8 & 0.875 & 855 & 931 \\ \hline
Early & 22.5 & 0.829 & 798 & 884 \\ \hline
Early + Late & 14 & 0.904 & 884 & 950 \\ \hline
\end{tabular}
\label{tab:Covers1000}
\end{table}

\begin{figure}
	\centering
	\includegraphics[width=0.8\textwidth]{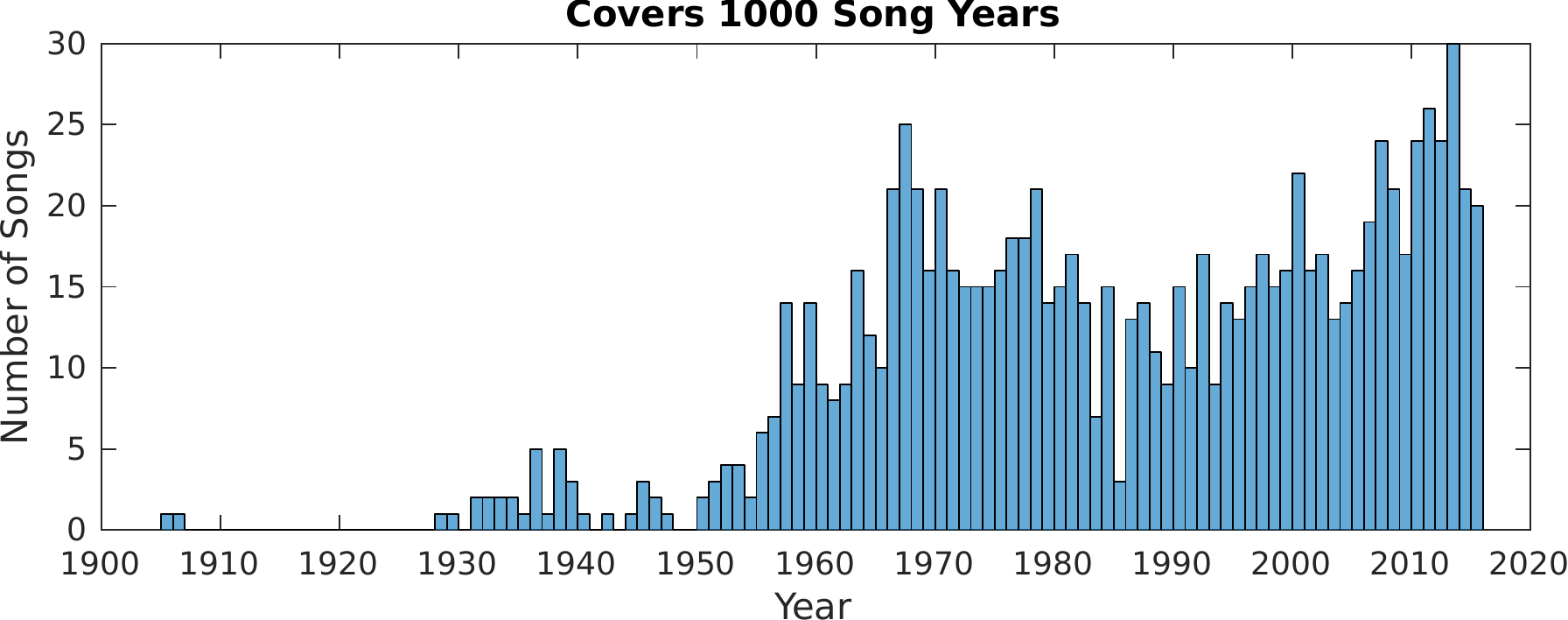}
	\caption{A distribution of years of songs in the Covers 1000 dataset.}
	\label{fig:Covers1000Years}
\end{figure}

To test our algorithm more thoroughly, we created our own dataset by manually choosing 1000 cover songs (395 cliques total) based on annotations given by users on \url{http://www.secondhandsongs.com}\footnote{MFCC and HPCP features for our dataset are publicly available at \url{http://www.covers1000.net}, along with beat intervals and other metadata including song title, album, and year}.  This dataset covers over a century of Western music from 1905 - 2016, and hence, it covers a wide variety of genres and styles.  Figure~\ref{fig:Covers1000Years} shows the full distribution of years covered.  By contrast, the Covers80 dataset contains almost exclusively pop music from the `80s and early `90s.

Most cliques have only two songs as in the Covers80 dataset, but there are a few cliques with 3 and 4 songs.  In this case, we report the MR and MRR of the first correctly identified song in the clique.  Table~\ref{tab:Covers1000} shows the results.  Similar trends are seen to the Covers80 case, and performance scales to this larger size.  One difference is that late fusion on HPCPs/MFCCs/MFCC SSMs performed better relative to early fusion, likely because the network was much richer with the additional volume of songs.

\subsection{Frank Zappa: ``The Black Page''}
\label{sec:zappa}

In our final experiment, we test a clique of 8 cover versions of the song ``The Black Page'' by Frank Zappa, which is entirely a drum solo that has absolutely no harmonic content.  We query each song against all of the songs in the Covers1000 dataset, and we compute the mean average precision (MAP) for the songs in the clique.  Unsurprisingly, for HPCP, the MAP is a mere 0.014, while for the rest of the features these songs are quite distinct from the rest of the songs in the Covers 1000 dataset.  The best performing feature set is early SNF, with a MAP of 0.98, followed by raw blocked MFCCs at a MAP of 0.97, followed by MFCC SSMs with a MAP of 0.905.

\section{Discussion}

In this work, we have demonstrated the benefit of combining complementary features at a very early stage of the cover song identification pipeline, in addition to the late fusion techniques in \cite{Chen2017CSFusion}.  Unlike \cite{Chen2017CSFusion} and other techniques, our algorithm works on a pair of songs and does not need a network of songs to improve performance, though we show that incorporating information from a network of songs (``late fusion'') can further improve results.  We showed that HPCP and MFCC features capture complementary information and are able to boost performance substantially over either alone.  In the process, we also developed a novel cross-similarity fusion scheme which was validated on several datasets, and which we believe could be useful beyond cover song identification in music structure analysis.

The main drawback of our technique is the requirement of beat tracking.  In practice, beat trackers may not return correct onsets.  Our current best remedy for this is to use different tempo biases, which blows up computation by a factor of 9.  Also, coming up with a single beat level is ill-posed, since most music consists of a hierarchy of rhythmic subdivisions \cite{quinton2015extraction}.  There does seem to be a recent convergence of techniques for rhythm analysis, though, \cite{degara2012reliability, krebs2015efficient} so hopefully our system will benefit.

In addition to imperfect beat intervals, there are also computational drawbacks in low level alignment, which is why most recent works on cover songs perform approximations to global cross-correlation, such as 2D Fourier Magnitude Coefficients \cite{ellis2012large, nieto2014music}.  By contrast, we rely on Smith Waterman, which is a quadratic algorithm, and early SNF adds another quadratic time complexity algorithm even with sparse nearest neighbor sets.  To address this, we are in the process of implementing GPU algorithms for every step of our pipeline, and we hope to apply it to the ``Second Hand Songs Dataset,'' which is a subset of the Million Songs Dataset \cite{bertin2011million}.

\section{Acknowledgements}
Christopher Tralie was partially supported by an NSF Graduate Fellowship NSF under grant DGF-1106401 and an NSF big data grant DKA-1447491.  We would also like to thank Erling Wold for pointing out the 8 covers of ``The Black Page'' by Frank Zappa, and we would like to thank the community at \url{www.secondhandsongs.com} for meticulously annotating songs which helped us to design Covers 1000.

\bibliographystyle{plain}
\bibliography{main}

\end{document}